# Web Video Categorization based on Wikipedia Categories and Content-Duplicated Open Resources

Zhineng Chen[1,2], Juan Cao[1], Yicheng Song[1,2], Yongdong Zhang[1], Jintao Li[1]
[1]Institute of Computing Technology, Chinese Academy of Sciences, Beijing 100190, China
[2]Graduate School of the Chinese Academy of Sciences, Beijing 100039, China
{chenzhineng, caojuan, songyicheng, zhyd, jtli}@ict.ac.cn

## ABSTRACT
This paper presents a novel approach for web video categorization by leveraging Wikipedia categories (WikiCs) and open resources describing the same content as the video, i.e., *content-duplicated open resources* (CDORs). Note that current approaches only collect CDORs within one or a few media forms and ignore CDORs of other forms. We explore all these resources by utilizing WikiCs and commercial search engines. Given a web video, its discriminative Wikipedia concepts are first identified and classified. Then a textual query is constructed and from which CDORs are collected. Based on these CDORs, we propose to categorize web videos in the space spanned by WikiCs rather than that spanned by raw tags. Experimental results demonstrate the effectiveness of both the proposed CDOR collection method and the WikiC voting categorization algorithm. In addition, the categorization model built based on both WikiCs and CDORs achieves better performance compared with the models built based on only one of them as well as state-of-the-art approach.

## Categories and Subject Descriptors
H.3.5 [**Information Storage and Retrieval**]: Web-based services

## 1. INTRODUCTION
Nowadays, we are experiencing an era of web information explosion: an event of interest is usually repeatedly recorded by web resources from various sites with different media forms, e.g., text document, video, image, etc. These resources construct an exhaustive set of information for the event and we denote them as *content-duplicated open resources* (CDORs). The CDORs, if appropriately collected and synthesized, are expected to be useful in many multimedia applications.

In this paper, we will investigate how to use external know-ledge bases, i.e., Wikipedia, to collect CDORs of a web video and how to use them to categorize web videos to the most suitable categories. As for a web video, typically, its CDORs include similar or duplicate videos and images, text documents describing the same content as the video, etc. To our knowledge, little work focuses on this topic although there are studies exploring open resources to improve the performance of video related applications.

*This work was supported by the National Basic Research Program of China (973Program, 2007CB311100), National Nature Science Foundation of China (60902090), Co-building Program of Beijing Municipal Education Commission

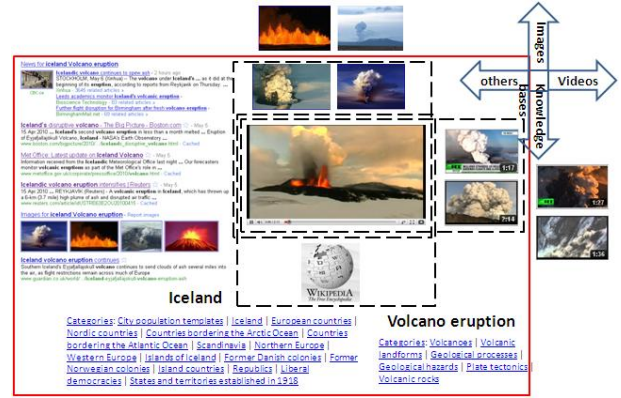

**Figure 1: Example of CDORs collection. Existing methods are only able to find CDORs of one or a few media forms. However, CDORs are of various media forms.**

Open resources are shown to be useful in many video research tasks. In recent TRECVID campaign, several teams [1-2] explore the use of external images to alleviate the scarcity of positive samples in the high-level feature extraction task, or to enrich expansion search strategies in the interactive search task. Observing that web video sharing sites (e.g.,YouTube) often have a amount of content redundancy, search and mining based video annotation methods are proposed in [3-5]. Wikipedia, the largest online encyclopedia, are also shown to be helpful in understanding user's intent [6-7]. In [8], Wu et al. propose a contextual and social combined web video categorization method. Their performance is significantly superior to the textual baseline, indicating the importance of open resources to the web video categorization task.

Fig.1 gives an illustrative summarization of the use of open resources in these works from the point of CDORs. As can be seen, given a video about the volcano eruption in Iceland, existing methods are only able to collect limited CDORs of one or a few media forms (the resources inside black dashed blocks), while many CDORs of other forms are totally ignored. If all these CDORs, i.e., the resources inside the red solid block, could be collected, then more knowledge about the video is known and a better video categorization result can be expected.

We believe, the reason that previous work is only able to explore CDORs within one or a few media forms, lies in that they do not seek CDORs by text ways. Now textual search engines are primary tools for people finding their interested resources regardless of media forms. In this paper we devote to studying the collection of multi-form CDORs with respect to a web video and the design of effective web video categorization algorithms on top of them by leveraging Wikipedia categories (WikiCs) and commercial search engines. Specifically, we employ WikiCs to identify and to classify discriminative concepts from title and tags so as to construct a query for a given video. By doing this, CDORs of various forms can be collected by issuing the query to Google search engine. Based on these CDORs, we constructively propose to implement the web video categorization task by WikiC features instead of conventional textual or visual features, where a WikiC voting algorithm is proposed. Experimental results show that both the proposed CDOR collection method and the WikiC voting algorithm are effective. In addition, we show that the categorization model built based on both WikiCs and CDORs achieves better performance compared with the models built based on either of them as well as state-of-the-art approach.

## 2. CDORs COLLECTION

Given a web video, CDORs can be collected in different ways, e.g., using visual similarities [1-4] or social proximities [8]. Here we collect CDORs by exploring WikiCs and commercial search engines. The method is a novel scheme and its main challenge is how to accurately capture the core content of the video.

### 2.1 Discriminative Concept Identification

Given a video, its discriminative concepts are identified by Wikipedia. Wikipedia has more than 3 million concepts, which build a complete dictionary containing almost all words and phrases. These concepts are characterized by hundreds of thousands of WikiCs created by human editors. It is observed that there are some deterministic relation between Wikipedia concepts and WikiCs. For example, a real person recorded by Wikipedia has a WikiC analogous to "1945 births", which records the birth year of the person. Similarly, a character in literatures, e.g., Harry Potter, has a WikiC "Harry Potter characters", indicating he is a virtual character in novel "Harry Potter". Therefore, terms "births" and "characters" are simple yet effective terms in recognizing whether a Wikipedia concept corresponds to a person. Based on these observations, we utilize Wikipedia to recognize the discriminative concepts among title and tags as well as their categories as follows.

First, we map title/tags to Wikipedia concepts according to the *Wikipedia longest match principle*. That is, for example, given two successive words "Expo", "2010". Both "Expo" and "Expo 2010" are detected as Wikipedia concepts, but only "Expo 2010" is reserved. The mapping makes sense because longer Wikipedia concepts are more likely to be specific and thus discriminative.

Second, we design a heuristic to classify concepts to person, location and other proper nouns according to whether certain representative terms (listed in Table.1) appear in their WikiCs. To differentiate concepts within the same class, we assume that concepts with more words or WikiCs are more important. Consequently, "George Bush" is regarded as more discriminative than "Bush" in class "person". Meanwhile, "George Bush" is also ranked higher than "Nouri Maliki" (Prime Minister of Iraq) as the number of WikiCs of Bush is larger than that of Maliki (38 vs 11).

A detail worth mentioning is the handling of ambiguous concepts, e.g., apple. When this happens, concept disambiguation is performed by measuring the WikiCs overlapping between each meaning of the concept and other tags associated with the video. The meaning overlapped most is treated as the correct meaning.

Note that Ref. [6, 10] also concern Wikipedia-based tag classification. However, their implementation is quite different with ours. In [6], the authors mainly use links between Wikipedia concepts to perform the classification while ours using WikiCs. As for [10], it categorizes tags to 11 WordNet noun categories. Compared with the two approaches, ours is much simple and achieves nearly perfect performance for person and location classes, and acceptable performance for other proper nouns class, according to our observations in the experiment.

### 2.2 CDORs Determination

We use two steps, i.e., query construction and resources search, to collect CDORs. The first step aims at choosing several discriminative concepts to construct a query that captures the core content of the video. The second step aims at gathering CDORs by submitting the query to Google search engine.

In the query construction, a basic problem is what kinds of Wikipedia concepts and how many of them are necessary to basically describe the core content of a video. To answer this question, we check 832 videos pertaining to the 73 hot topics discovered from MCG-WEBV [9]. We manually extract core concepts from title and tags for each video. Our finding is that 2.7 Wikipedia concepts, with 0.7 person, 0.4 location, and 1.6 other proper nouns, are necessary for each video in average. Therefore, we pick 3 concepts, i.e., the most discriminative person, location and other proper noun, to construct the query. Note that person or location concepts are missing for many videos, and in such cases, we add other proper nouns to meet the number requirement.

We submit the query to Google search engine, where the top 20 returned results are collected. We evaluate the results one by one from back to forward. The process is stopped when all query concepts simultaneously appear in Google generated title or abstract of a result. The result and results ranked above are regarded as CDORs. For the case where the number of CDORs is less than 5, the top 5 results are regarded as CDORs by default.

As expected, the CDORs are of various forms including video, image, news and blog article, etc, which supply rich and complement knowledge about the content of the video. Considering the difficulty of crawling complete version of these CDORs for users that are external to search engines, we only collect the Google generated titles and abstracts. Nevertheless, these textual descriptions still supply much richer knowledge than raw tags and their effectiveness is demonstrated in the experiment.

**Table 1. Representative terms in the proposed heuristic**

| Class | Representative terms |
|---|---|
| person | births, characters |
| location | cities, countries, geography of, states of, regions of, provinces of, museums in, landmarks in, capitals in, islands of, boroughs of, stadiums, airports, locations, geography stubs, places |
| other proper nouns | companies, vehicles, devices, established in, games, establishments, venues, inc., songs, films, services, television series, websites, cartoons, books, incidents novels, albums, agents, teams, brands, cameras, shows, magazines, awards, graphics, inventions, drugs, sports, introductions, genres, occupations, foods, articles |

## 3. WIKIC VOTING CATEGORIZATION

In this section, we investigate the use of WikiCs to categorize web videos. This is reasonable as: 1) WikiCs organize concepts in accordance with human cognition, which may be more useful than raw concepts in the categorization; 2) the limited concepts are greatly enriched when mapping to the WikiC space, as a Wikipedia concept always corresponds to several to dozens of WikiCs.

Therefore, we perform the web video categorization task as follows. First, we design an *enriched WikiC* (EWikiC) expression on top of the raw one, where inductive terms in WikiCs, like "Capitals" and "Asia" in WikiC "Capitals in Asia", are extracted and added. EWikiCs provide more valuable clues to predict the category label of a video and it performs superior to the model merely using WikiCs in our test.

Then, we design a WikiC voting algorithm based on EWikiCs to implement the categorization. Denote by $\text{Cate}_k$ the predicted category label of video $V_k$, it is determined by the category with the highest predicted value as follows.

$$\text{Cate}_k = \arg\max_{i=1}^{n}\left(\sum_{j \in D_k} p(c_i|f_j)\right) \quad (1)$$

where $n$ is the number of predefined video categories, $D_k$ is the set of Wikipedia concepts of $V_k$. $p(c_i|f_j)$ is the conditional probability of video category $i$ given concept $j$ and it is computed by

$$p(c_i|f_j) = \sum_{t \in \mathcal{H}_j} Vote(c_i|w_t) \quad (2)$$

where $\mathcal{H}_j$ is the set of EWikiCs of concept $j$, $Vote(c_i|w_t)$ is the confidence voting from EWikiC $t$ to video category $i$ and it is given by

$$Vote(c_i|w_t) = \frac{1}{E(w_t)} \cdot \frac{\varphi(w_t|c_i)}{\sum_{i=1}^{n}\varphi(w_t|c_i)} \quad (3)$$

In Eq. 3, $E(w_t)$ is the entropy of EWikiC $t$ among the $n$ video categories and it is defined as

$$E(w_t) = -\sum_{i=1}^{n} \tilde{\varphi}(w_t|c_i) \cdot \log_2(\tilde{\varphi}(w_t|c_i)) \quad (4)$$

where $\tilde{\varphi}(w_t|c_i)$ is the normalized $\varphi(w_t|c_i)$. $\varphi(w_t|c_i)$ is a value indicating the proximity of EWikiC $t$ and video category $i$. It is estimated by

$$\varphi(w_t|c_i) = \sum_{k \in TrainSet} \frac{Fre_k(w_t|c_i)}{\sum_{t \in S_k} Fre_k(w_t|c_i)} \quad (5)$$

where $TrainSet$ is the set of training videos whose category labels are given, $S_k$ is the set of EWikiCs belonging to the concepts associated with $V_k$. $Fre_k(w_t|c_i)$ is the frequency that EWikiC $t$ appears in $S_k$ if the label of $V_k$ is $i$, and 0 otherwise.

By using the above voting scheme, WikiC based categorization models are built based on EWikiCs, which decides category labels of web videos mainly by EWikiCs distribution over the categories.

## 4. EXPERIMENTS
### 4.1 Dataset description

To demonstrate the effectiveness of both the CDORs collection method and the WikiC voting algorithm, we perform experiments on MCG-WEB [9]. The dataset contains two parts[1], with 80,031 YouTube videos crawled from Dec 2008 to Feb 2009 in part-1, and 5887 "Most Viewed" YouTube videos of "This month" from Mar. 2009 to Jul. 2009 in part-2. 10,000 videos randomly selected from part-1 are used as training set and the whole part-2 are used as testing set. For both training and testing sets, we identify Wikipedia concepts from their titles and tags. There are 35,691 unique concepts and 93,236 unique EWikiCs, which are defined as raw concepts and raw EWikiCs, respectively. In addition, the concepts cover 96.3% of the tags that appear more than 5 times.

---
[1] The dataset has been updated to a larger scale and can be accessed at http://mcg.ict.ac.cn/mcg-webv.htm.

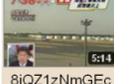

**Figure 2: Typical queries and collected Wikipedia concepts.**

### 4.2 The quality of the collected CDORs

We evaluate the quality of collected CDORs in terms of both coverage of the extracted concepts and relevance of the collected CDORs. For the first measurement, we compare the query concepts extracted by the proposed method with the human labeled ones among the 873 videos. The query concepts cover 83.1% of the labeled concepts, indicating that for most videos, the proposed method can construct queries as accurate as human.

To evaluate the relevance, we construct queries for 15,453 of the 15,887 videos, from which 307,209 open resources are found and 232,952 of them are judged as CDORs. We further identify Wikipedia concepts in their Google generated titles and abstracts, which build an enriched concept set and an enriched EWikiC set. Three typical examples are listed in Fig. 2. As can be seen, the constructed queries generally capture the core content of the videos and many relevant concepts are collected accordingly, indicating that the resources collected by Google are indeed CDORs.

### 4.3 Categorization performance

We carry out web video categorization experiment on both raw concepts and enriched concepts to further demonstrate the effectiveness of the proposed CDORs collection method and the WikiC voting algorithm. SVM classifiers based on "bag-of-word" features (i.e., concept vector) are employed for comparison. Their combinations lead to four schemes, i.e., SVM on raw concepts (RC-SVM), SVM on enriched concepts (EC-SVM), WikiC voting on raw concepts (RC-Wiki), and WikiC voting on enriched concepts (EC-Wiki). For all the four schemes, models are first trained on the training data and then predicted on the testing data. The YouTube provided 15 categories and the raw video category label are used as the ground truth. The mean average precision (MAP) is adopted as the performance metric. Experimental results of all the four schemes are shown in Fig.3.

#### 4.3.1 The effectiveness of CDORs

It is seen from Fig.3 that steady improvements are obtained by incorporating CDORs (a 9.4% improvement for SVM classifier and an 11.2% improvement for WikiC Voting). We argue the gain in EC-SVM mainly attributes to missing but valuable concepts are discovered from CDORs, therefore the sparse concept vector is enriched. On the other hand, the enriched concepts also provide constructive clues for WikiC voting, as they describe the same video and thus are very likely to share or have similar EWikiCs. Therefore the discriminability of the WikiC voting models is reinforced and gives rise to improvements.

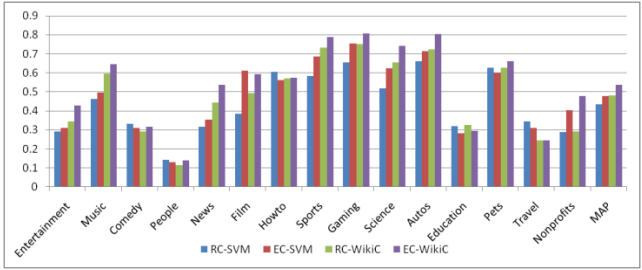

**Figure 3: Web video categorization performance for RC-SVM, EC-SVM, RC-Wiki, and EC-Wiki.**

### 4.3.2 The effectiveness of WikiC Voting

It is also seen from Fig.3 that the WikiC voting performs better than its SVM counterpart on both raw and enriched concepts (10.2% and 12.7% higher on raw and enriched concepts). This is reasonable as tens of thousands EWikiCs are used in the WikiC voting while its SVM counterpart only uses nearly 4500 concepts due to computational costs. Another reason the WikiC voting is superior to SVM is that EWikiCs reflect human cognition to some extent and thus more informative than raw concepts. For example, from the perspective of WikiCs, "Donald Duck" and "Mickey Mouse" are very similar as they have dozens of EWikiCs in common, but the relation may be difficult to learn by a concept vector classifier as they have no common words.

### 4.3.1 The overall performance

The performance of EC-Wiki is 23.3% better than that of RC-SVM. It is significant and is attributed to using CDORs and performing the categorization task EWikiC. For well-defined categories such as "Gaming", "Science&Technology", "Pets&Animals", "Music", "Sports", and "Autos&Vehicles", RC-SVM shows decent performance but EC-Wiki can further improve it, indicating "rules of thumb" of these categories is successfully captured. On the other hand, for categories "People&Blogs" "Travel&Events", "Comedy" and "Education", the performance of the baseline is unsatisfactory and EC-Wiki do not show its superiorityeither, indicating limited "rules of thumb" is learned. We argue that is because the videos belonging to these categories are general and diverse, which makes these categories less specific.

Note that for web video categorization, Wu et al [8] get state-of-the-art performance by combining textual classification, related video and author voting. We implement our method by using EC-Wiki instead of their textual classification, and compare with their method in our dataset. The experimental results are given in Table.2. Although quite good performance is achieved by their method, ours is still able to improve the performance by 3.4%, which further illustrates the superiority of the proposed method.

## 5. CONCLUSION

In this paper, we investigate the method of collecting and using CDORs and WikiCs to improve the performance of web video categorization. There are two major contributions. First, we propose a method that bridges the understanding of web video tags by means of WikiCs and collects multi-form CDORs through Google search engine. Second, we introduce a voting algorithm that categorizes web video in the space spanned by EWikiCs rather than that spanned by raw title and tags. Experimental results show improvements obtained by the incorporation of CDORs and EWikiCs are as high as 23.3%. We believe our work will provide valuable insights in the use of open resources to benefit web video related applications.

**Table 2. Performance of 15 categories for related video voting (R-Voting), author voting (A-Voting), Wu's and ours.**

|  | *R-Voting* | *A-Voting* | *Wu's* [8] | *Ours* |
|---|---|---|---|---|
| Entertainment | 0.566587 | 0.651354 | 0.731229 | **0.749912** |
| Music | 0.616263 | 0.68958 | 0.801728 | **0.80781** |
| Comedy | 0.882759 | 0.817609 | 0.90434 | **0.911331** |
| People | 0.433709 | 0.491006 | **0.605631** | 0.583107 |
| News | 0.580541 | 0.578184 | **0.705651** | 0.701221 |
| Film | 0.681421 | 0.526861 | 0.800813 | **0.804712** |
| Howto | 0.898057 | 0.896468 | 0.907756 | **0.938459** |
| Sports | 0.820215 | 0.656078 | 0.871721 | **0.894002** |
| Gaming | 0.909265 | 0.867428 | 0.93069 | **0.948561** |
| Science | 0.836621 | 0.78787 | 0.861061 | **0.886475** |
| Autos | 0.915815 | 0.886108 | 0.940603 | **0.961684** |
| Education | 0.643414 | 0.682143 | 0.66462 | **0.766708** |
| Pets | 0.85638 | 0.815959 | 0.866783 | **0.909485** |
| Travel | 0.686221 | 0.722041 | 0.723593 | **0.788834** |
| Nonprofits | 0.755539 | 0.797747 | 0.785031 | **0.85886** |
| MAP | 0.738854 | 0.724429 | 0.80675 | **0.834077** |